\documentclass[letterpaper,twocolumn,prl,aps,superscriptaddress,amsmath,amssymb,floatfix]{revtex4-1}
\usepackage{mathptmx}
\DeclareMathAlphabet{\mathcal}{OMS}{cmsy}{m}{n}
\usepackage[latin9]{inputenc}
\usepackage{color}
\usepackage{verbatim}
\usepackage{float}
\usepackage{amsmath}
\usepackage{amssymb}
\usepackage{graphicx}
\usepackage{esint}
\usepackage[unicode=true,
 bookmarks=true,bookmarksnumbered=false,bookmarksopen=false,
 breaklinks=false,pdfborder={0 0 1},backref=false,colorlinks=true]
 {hyperref}
\hypersetup{
 linkcolor=magenta,urlcolor=blue,citecolor=blue,pdfstartview={FitH},hyperfootnotes=false}

\makeatletter

\usepackage{textcomp}
\usepackage{epstopdf}

\pdfpageheight\paperheight
\pdfpagewidth\paperwidth

\@ifundefined{textcolor}{}{%
 \definecolor{BLACK}{gray}{0}
 \definecolor{WHITE}{gray}{1}
 \definecolor{RED}{rgb}{1,0,0}
 \definecolor{GREEN}{rgb}{0,1,0}
 \definecolor{BLUE}{rgb}{0,0,1}
 \definecolor{CYAN}{cmyk}{1,0,0,0}
 \definecolor{MAGENTA}{cmyk}{0,1,0,0}
 \definecolor{YELLOW}{cmyk}{0,0,1,0}
}

\usepackage{xcolor}
\usepackage{soul}
\newcommand{\bra}[1]{\ensuremath{\left\langle#1\right|}}
\newcommand{\ket}[1]{\ensuremath{\left|#1\right\rangle}}

\definecolor{blue}{rgb}{0,0,1}
\definecolor{red}{rgb}{1,0,0}
\definecolor{green}{rgb}{0,1,0}

\@ifundefined{textcolor}{}{%
 \definecolor{BLACK}{gray}{0}
 \definecolor{WHITE}{gray}{1}
 \definecolor{RED}{rgb}{1,0,0}
 \definecolor{GREEN}{rgb}{0,1,0}
 \definecolor{BLUE}{rgb}{0,0,1}
 \definecolor{CYAN}{cmyk}{1,0,0,0}
 \definecolor{MAGENTA}{cmyk}{0,1,0,0}
 \definecolor{YELLOW}{cmyk}{0,0,1,0}
}

\usepackage{xcolor}\usepackage{soul}
\newcommand{\abs}[1]{\left| #1 \right|}

\definecolor{blue}{rgb}{0,0,1}
\definecolor{red}{rgb}{1,0,0}
\definecolor{green}{rgb}{0,1,0}

\usepackage{soul}

\makeatother

\begin{document}
\title{Efficient implementation of arbitrary Hermitian-preserving and trace-preserving maps}

\affiliation{Laboratory of Quantum Information, University of Science and Technology of China, Hefei 230026, China}

\affiliation{QICI Quantum Information and Computation Initiative, School of Computing and Data Science, The University of Hong Kong, Pokfulam Road, Hong Kong}

\affiliation{Thrust of Artificial Intelligence, Information Hub, The Hong Kong University of Science and Technology (Guangzhou), Guangdong 511453, China}

\affiliation{Center for Quantum Information, Institute for Interdisciplinary Information Sciences, Tsinghua University, Beijing 100084, China}

\affiliation{Hefei National Laboratory, Hefei 230088, China}

\author{Weizhou~Cai}

\affiliation{Laboratory of Quantum Information, University of Science and Technology of China, Hefei 230026, China}

\author{Zi-Jie~Chen}

\affiliation{Laboratory of Quantum Information, University of Science and Technology of China, Hefei 230026, China}

\author{Xuanqiang~Zhao}

\affiliation{QICI Quantum Information and Computation Initiative, School of Computing and Data Science, The University of Hong Kong, Pokfulam Road, Hong Kong}

\author{Xin Wang}
\affiliation{Thrust of Artificial Intelligence, Information Hub, The Hong Kong University of Science and Technology (Guangzhou), Guangdong 511453, China}

\author{Guang-Can~Guo}
\affiliation{Laboratory of Quantum Information, University of Science and Technology of China, Hefei 230026, China}
\affiliation{Hefei National Laboratory, Hefei 230088, China}

\author{Luyan~Sun}
\email{luyansun@tsinghua.edu.cn}
\affiliation{Center for Quantum Information, Institute for Interdisciplinary Information
Sciences, Tsinghua University, Beijing 100084, China}
\affiliation{Hefei National Laboratory, Hefei 230088, China}

\author{Chang-Ling~Zou}
\email{clzou321@ustc.edu.cn}
\affiliation{Laboratory of Quantum Information, University of Science and Technology of China, Hefei 230026, China}
\affiliation{Hefei National Laboratory, Hefei 230088, China}

% \date{\today}

\begin{abstract}
Quantum control has been a cornerstone of quantum information science, driving major advances in quantum computing, quantum communication, and quantum sensing. Over the years, it has enabled the implementation of arbitrary completely positive and trace-preserving (CPTP) maps; an important next step is to extend control to Hermitian-preserving and trace-preserving (HPTP) maps, which underpin applications such as entanglement detection, quantum error mitigation, quantum simulation, and quantum machine learning. Here we present an efficient and fully constructive method for implementing arbitrary HPTP maps. Unlike existing methods that decompose an HPTP map into multiple CPTP maps or approximate it using bipartite Hamiltonians with large Hilbert spaces, our approach compiles a target HPTP map into a single executable CPTP map whose Kraus rank is guaranteed to be no larger than the intrinsic rank of the target HPTP map plus one, followed by simple classical post-processing.
Numerical results for inverse noise channels used in quantum error mitigation, including bosonic photon loss, confirm substantial reductions in resources and highlight scalability in higher-dimensional settings.
Together with our numerical benchmarks, these results validate the efficiency and versatility of the proposed framework, opening a route to broader quantum-information applications enabled by HPTP processing.
\end{abstract}

%\date{\today}

\maketitle

\begin{figure}
\centering{}\includegraphics[width=\columnwidth]{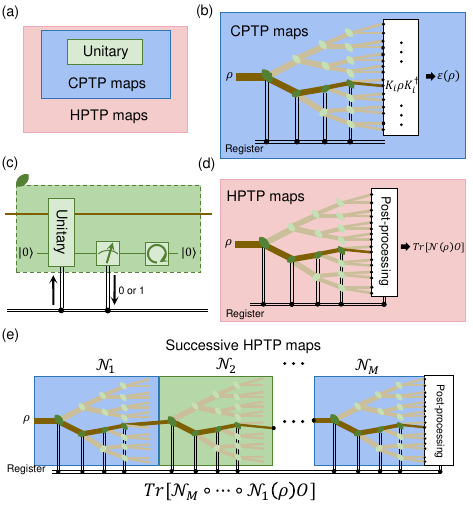}
\caption{Concept for HPTP maps and their realization. (a) Hierarchy of unitary operations, CPTP maps, and HPTP maps. (b) Binary tree-structured quantum circuit for implementing arbitrary CPTP maps. (c) Quantum circuit for each leaf of the binary tree-structured circuit, consisting a joint unitary operation on a composed quantum system, a projective measurement on the ancilla qubit, and a resetting process for the ancilla qubit. (d) Binary tree-structured quantum circuit for implementing arbitrary HPTP maps. (e) Binary tree-structured quantum circuit for implementing successive HPTP maps.}
\label{fig1}
\end{figure}

%\section{Introduction}
\noindent\textit{Introduction.-} Over the past few decades, quantum information science has made remarkable progress, leading to significant advancements in quantum computing, quantum communication, and quantum sensing. These rapid developments have been underpinned by sophisticated quantum control techniques, enabling researchers to manipulate the quantum states of physical systems. Figure~\ref{fig1}(a) illustrates the hierarchy of quantum control technology. Early efforts focused on unitary operations on closed quantum systems, including preparation of states~\cite{Law1996,Hofheinz2009} and implementation of unitary gates~\cite{Khaneja2005,Dawson2005,Krastanov2015,Heeres2017}. However, truly isolated quantum systems do not exist, as all systems inevitably interact with environments, which motivates the second level of the hierarchy: completely positive and trace-preserving (CPTP) maps, which describe the most general physically realizable quantum processes.
With the ability to implement arbitrary CPTP maps, one can generate entangled states from mixed states~\cite{Barreiro2011An}, simulate open quantum system dynamics~\cite{Lu2017,Hu2018channel,Han2021}, and implement generalized measurements~\cite{Cai2021PRL,Cai2024}.

Beyond CPTP maps, an important frontier in quantum control is to realize maps that are not physical quantum channels, such as the partial transpose used for entanglement detection~\cite{Peres1996,Horodecki2009}, inverse noise channels central to quantum error mitigation (QEM)~\cite{Temme2017PRL,CaiQEM}, and non-CP intermediate maps that can arise in non-Markovian dynamics and can also be useful in quantum machine learning as flexible feature-extraction primitives~\cite{Cong2019}. All these non-physical maps are non-CP, and their combination with CPTP contributes to more complete description as Hermitian-preserving and trace-preserving (HPTP) maps~\cite{Regula2021,Jiang2021Quantum,Cao2023,Wei2024}. Extending the quantum control for realizing arbitrary HPTP maps beyond the CPTP is appealing for a broad class of quantum technologies. Recently, several approaches have been proposed to bridge this gap. One method decomposes the target HPTP map into two CPTP maps and post-processing the measurement results after implementing both CPTP maps~\cite{Regula2021,Jiang2021Quantum}. This method requires semi-definite programming to optimize the implementation and reduce the sampling overhead, and its experimental demonstration has been limited to single-qubit systems~\cite{Rossini2023PRL}. Another approach constructs a bipartite Hamiltonian  equal to the partial transpose of the Choi matrix to approximate the target HPTP map~\cite{Wei2024}, and thus requires an ancilla of the same dimension as the target system.

In this Letter, we present a fully constructive and resource-efficient method to realize arbitrary HPTP maps using a single executable CPTP map followed by simple classical post-processing. The Kraus rank is guaranteed to be no larger than the intrinsic rank $r$ of the target HPTP map plus one, and allows binary tree-structured circuit implementation with a circuit depth of $\mathrm{log}_2(r+1)$ that requires only a single two-level ancilla. Compared with previous approaches, our protocol substantially reduces quantum resources and circuit depth, while remaining directly compatible with near-term hardware. Additionally, our method provides a transparent, analytically guaranteed sampling cost. We further characterize the required Kraus rank and the sampling (variance) cost for estimating expectation values, and validate the advantages of our method numerically on representative inverse noise channels relevant to QEM, including bosonic photon loss.

\smallskip{}
\noindent\textit{Binary-tree Construction for CPTP maps.-} Since all physically realizable quantum processes are CPTP, the ability to implement arbitrary CPTP maps provides a universal toolbox for controlling open system dynamics.
Any CPTP map $\mathcal{E}$ admits a Kraus representation $\mathcal{E}(\rho)=\Sigma_i^r K_i \rho K_i^{\dagger}$, where $\rho$ represents a density matrix and $r$ is the Kraus rank, i.e., the minimal number of Kraus operators required in an operator-sum form~\cite{Nielsen}. A generic unitary dilation realizes $\mathcal{E}$ by coupling the system to ancillas and applying a joint unitary, but the required ancilla dimension can be costly~\cite{Nielsen}. To reduce quantum resources, feedback-assisted constructions have been developed. For a single qubit, Ref.~\cite{Wang2013PRL} showed that an arbitrary CPTP map can be realized using only one ancilla qubit and measurement-based feedback, by expressing the target channel as a convex combination of two rank-2 CPTP maps. This approach has been experimentally demonstrated in a photonic qubit of superconducting quantum systems~\cite{Hu2018channel}. For general $d$-dimensional systems, Ref.~\cite{Shen2017PRB} introduced a binary-tree compilation that implements an arbitrary CPTP map of Kraus rank $r$ using a single ancilla qubit that is repeatedly measured, reset, and reused, together with a classical register storing the measurement outcomes, as sketched in Fig.~\ref{fig1}(b). Each leaf node in this diagram corresponds to an implementation of the basic experimental quantum circuit module shown in Fig.~\ref{fig1}(c).
Operationally, each run of the circuit stochastically selects a Kraus branch and outputs the state proportional to $K_i\rho K_i^{\dagger}$ with probability $\mathrm{Tr}[K_i\rho K_i^{\dagger}]$, thereby realizing the target CPTP map upon averaging over runs.

The binary-tree structure provides an efficient scaling with the Kraus rank that a circuit with $n$ layers can realize up to $2^n$ Kraus branches, so implementing a map with rank $r$ requires a depth $n=\lceil \mathrm{log}_2(r)\rceil$. In the worst case $r= d^2$, yielding a depth upper bound $\lceil2\mathrm{log}_2 d\rceil$. This approach has enabled experimental realizations of arbitrary CPTP maps on a high-dimensional quantum system in a superconducting microwave cavity~\cite{Cai2021PRL}, and will serve as a building block for our HPTP construction below.

%\section{Construction of high-efficiency arbitrary HPTP maps}
\smallskip{}
\noindent\textit{Efficient construction of arbitrary HPTP maps.-} HPTP maps encompass all CPTP maps, but the subset of HPTP maps that excludes CPTP maps preserves Hermiticity when applied to a density matrix, but does not guarantee positivity. The Choi matrix representation of an HPTP map $\mathcal{N}$ is given by $\Lambda/d=\mathcal{I}\otimes\mathcal{N}(\ket{\Phi}\bra{\Phi})$, where $d$ is the system dimension, $\mathcal{I}$ denotes the identity operation, and $\ket{\Phi}$ is a maximally entangled state. Unlike a CPTP map, whose Choi matrix is positive semidefinite, the Choi matrix of a general HPTP map is Hermitian but may possess negative eigenvalues. By diagonalizing $\Lambda$, the Kraus operator $K_i$ of $\mathcal{N}$ can be obtained by reshaping the vector $\sqrt{\abs{\lambda_i}} \ket{i}$ into a matrix, where $\ket{i}$ is an eigenvector of $\Lambda$ with eigenvalue $\lambda_i$. The resulting Kraus operator representation of $\mathcal{N}$ reads
\begin{equation}
    \mathcal{N}(\rho)=\Sigma_{i=1}^r \alpha(i) K_i\rho K_i^\dagger,
\end{equation}
where $\alpha(i)=1$ or $-1$, corresponding to the sign of the eigenvalue of $\Lambda$. The trace-preserving condition reads $\Sigma_{i=1}^r \alpha(i) K_i^\dagger K_i=I$. For a CPTP map $\mathcal{E}$, all $\alpha(i)=+1$ for all $i$, whereas negative $\alpha(i)$ indicates that $\mathcal{N}$ cannot be implemented as a physical channel with positive outcome probabilities. Practical realizations must therefore be indirect and are naturally formulated at the level of expectation value.

The central idea of our method is to construct a single CPTP map  $\mathcal{E}_{\mathcal{N}}$ corresponding to the target HPTP map $\mathcal{N}$ and implementing it using a binary tree-structured circuit, with the non-physical negative signs $\alpha(i)$ absorbed into classical post-processing. Figure~\ref{fig1}(d) presents the construction, consisting of three steps:
\begin{enumerate}
    \item \textit{Normalization}. To construct the map $\mathcal{E}_{\mathcal{N}}$, we first normalize the sum $\Sigma_{i=1}^r K_i^\dagger K_i$ by a factor $\gamma$, ensuring that its norm is 1. Since $\Sigma_{i=1}^r  K_i^\dagger K_i\geq I$, we define $\gamma$ as its norm (the largest eigenvalue), with $\gamma\geq1$. We then introduce a set of rescaled Kraus operators $\{\tilde{K_i}\}$ with $\tilde{K}_i=K_i/\sqrt{\gamma}$ and construct a new CP map.
    \item \textit{Completion}. To ensure the resulting map remains CPTP, we add one additional Kraus operator, $\tilde{K}_{r+1}=(I-\Sigma_{i=1}^{r}  \tilde{K}_i^\dagger \tilde{K}_i)^{\frac{1}{2}}$, which guarantees completeness $\Sigma_{i=1}^{r+1}  \tilde{K}_i^\dagger \tilde{K}_i=I$. Thus, we obtain a CPTP map $\mathcal{E}_{\mathcal{N}}$ with Kraus operators $\{\tilde{K}_i\}_1^{r+1}$, corresponding to the target HPTP map $\mathcal{N}$.
    \item \textit{Post-processing}. To compute the expectation value of an observable $O$ after applying $\mathcal{N}$ to a state $\rho$, we post-process measurement results on a classical computer, multiplying each term $\mathrm{Tr}[\tilde{K}_i\rho\tilde{K}_i^{\dagger}O]$ by $\alpha(i)\gamma$ and the term $\mathrm{Tr}[\tilde{K}_{r+1}\rho\tilde{K}_{r+1}^{\dagger}O]$ by zero. The index $i$ of each Kraus operator is recorded by a classical register, and the final expectation value is given by
    \begin{equation}
        \mathrm{Tr}[\mathcal{N}(\rho)O]=\gamma\Sigma_{i=1}^r\alpha(i) \mathrm{Tr}[\tilde{K}_i\rho\tilde{K}_i^{\dagger}O].
    \end{equation}
\end{enumerate}

This three-step procedure is fully constructive as it requires no numerical optimization, and directly yields a CPTP map with a Kraus rank of $r+1$ that is compatible with the binary tree circuit~\cite{Andersson2008,Shen2017PRB}. In particular, when $\Sigma_{i=1}^r  K_i^\dagger K_i= \gamma I$, i.e., the original Kraus operator set $\{\tilde{K_i}\}$ is already complete, then the Kraus rank of $\mathcal{E}_\mathcal{N}$ exactly matches that of $\mathcal{N}$ without requiring the additional operator ($\{\tilde{K}_i\}_1^{r+1}$). Therefore, our method requires at most $\lceil \mathrm{log}_2(r+1)\rceil$ circuit depth, making the implementation of general HPTP maps feasible.

The sampling overhead determines the number of quantum state preparations required to estimate the expectation value $\langle O\rangle =\mathrm{Tr}[\mathcal{N}(\rho)O]$ to a given precision. If $\mathcal{N}$ were directly implementable as a physical process, the single-shot variance would be
\begin{equation}
\mathrm{Var}(O)=\langle O^2\rangle -\langle O\rangle ^2
=\mathrm{Tr}[\Sigma_{i=1}^{r} K_i\rho K_i^{\dagger} O^2]-\langle O\rangle ^2.
\label{eq:variance}
\end{equation}
The expectation value in our approach is given by:
\begin{equation}
\begin{split}
\langle O\rangle =&\mathrm{Tr}[\mathcal{N}(\rho)O]\\
=&\gamma \Sigma_{i=1}^{r} \alpha(i)\mathrm{Tr}[\tilde{K}_i\rho \tilde{K}_i^{\dagger}O]+0\mathrm{Tr}[\tilde{K}_{r+1}\rho \tilde{K}_{r+1}^{\dagger}O].
\label{eq:expect}
\end{split}
\end{equation}
Based on the variance of a mixed quantum state with respect to an observable~\cite{Petruccione2002}, we can derive the single-shot variance of our method:
\begin{equation}
\begin{split}
\mathrm{Var}_{\text{our}}(O)=&\gamma^2\Sigma_{i=1}^r \mathrm{Tr}[\tilde{K}_i \rho \tilde{K}_i^{\dagger}O^2] -\langle O\rangle^2\\
=&\gamma\Sigma_{i=1}^r \mathrm{Tr}[K_i \rho K_i^{\dagger}O^2] -\langle O\rangle^2,
\label{eq:var_single}
\end{split}
\end{equation}
where the set of Kraus operators $\{\tilde{K}_i\}_{i=1} ^{r+1}$ belongs to the CPTP map $\mathcal{E}_{\mathcal{N}}$ which is actually applied to the state $\rho$. The measurement results corresponding to the $r+1$-th Kraus operator are multiplied by zero during post-processing.
In an experiment, the circuit is repeated $N$ times and the empirical mean $\frac{1}{N}\sum_{j=1}^{N}o_j$ is used to estimate $\langle O\rangle$, where $o_j$ is the reweighted outcome in the $j$-th run. The variance of this mean estimator is therefore:
\begin{equation}
\begin{split}
\mathrm{Var}_{\text{our}}(\frac{\Sigma_jo_j} {N})
=\frac{\gamma\Sigma_{i=1}^r \mathrm{Tr}[K_i \rho K_i^{\dagger}O^2] -\langle O\rangle^2}{N}.
\label{eq:var_N}
\end{split}
\end{equation}
Further derivation details are provided in the Supplementary Materials~\cite{SI}.

Beyond the practical advantages of our binary-tree circuit implementation, including two-level ancilla reuse, shallow circuit depth, and fully deterministic construction without numerical optimization, our analytical results of the variance provide a direct and transparent sampling overhead. Given a target HPTP map, an input state, and an observable, we can evaluate the number of repetitions required to achieve any desired estimation precision. This stands in sharp contrast to previous two-CPTP decomposition method~\cite{Regula2021,Jiang2021Quantum}, where $\mathcal{N}=\eta_1\mathcal{E}_1-\eta_2\mathcal{E}_2$, the sampling overhead is determined by coefficients $\eta_1$ and $\eta_2\geq0$ of two CPTP maps ($\mathcal{E}_{1,2}$) obtained through the semi-definite programming.

Due to the post-processing involved in our method, the output of the quantum circuit in Fig.~\ref{fig1}(d) is the expectation value $\mathrm{Tr}[\mathcal{N}(\rho)O]$ rather than a quantum state. As a result, the quantum state cannot be reused for subsequent quantum processes after the post-processing. To implement a sequence of successive HPTP maps, yielding $\mathrm{Tr}[\mathcal{N}_M\circ...\circ\mathcal{N}_1(\rho)O]$, we can concatenate quantum circuits corresponding to all target HPTP maps and perform a single post-processing step at the end, as shown in Fig.~\ref{fig1}(e), where the classical register restores the Kraus-branch indices (measurement outcomes of the ancilla) for each applied map and the final weight is determined from all recorded outcomes.

\begin{figure}
\centering{}\includegraphics[width=\columnwidth]{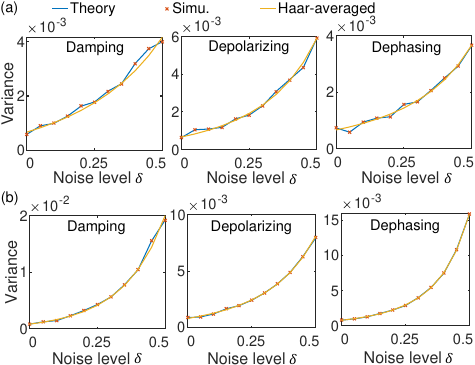}
\caption{Simulation of QEM through implementing HPTP maps. (a) Mitigation of quantum noise for a single qubit. (b) Mitigation of quantum noise for two qubits.}
\label{fig2}
\end{figure}

%\section{applications}
\smallskip{}
\noindent\textit{Applications.-} The ability to efficiently implement arbitrary HPTP maps opens pathways to various interesting quantum information applications. A prominent example is QEM, where a central task is recovering near-ideal expectation values from noisy quantum states. Probabilistic quantum error cancellation~\cite{Temme2017PRL}, a leading QEM technique, fundamentally relies on effectively implementing inverses of noise channels, corresponding to HPTP maps~\cite{Jiang2021Quantum}. By constructing an HPTP map $\mathcal{N}=\mathcal{E}^{-1}$ that inverts a given noise channel $\mathcal{E}$, we can mitigate the effects of noise and recover the ideal expectation value, satisfying $\mathrm{Tr}[\mathcal{N}\circ\mathcal{E}(\rho)O]=\mathrm{Tr}[\rho O]$. Here, we demonstrate the mitigation of three types of quantum noise using our approach.% and compare the results with the previous approach proposed in~\cite{Regula2021,Jiang2021Quantum}.

\begin{figure}
\centering{}\includegraphics[width=\columnwidth]{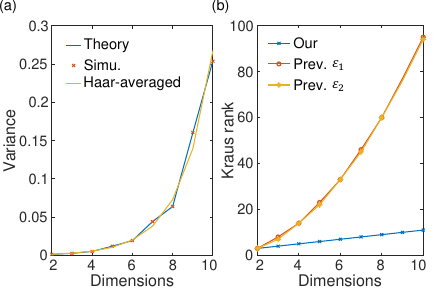}
\caption{Performance of QEM on a bosonic mode. (a) Variance in mitigating photon loss errors as a function of system dimension. (b) Kraus rank for mitigating photon loss errors as a function of system dimension. The error rate is fixed at $\delta=0.2$ for different system dimensions (see Supplementary Materials~\cite{SI} for the definition of the error rate).}
\label{fig3}
\end{figure}

Figure~\ref{fig2} illustrates the variance as a function of noise level for mitigating amplitude damping noise, depolarizing noise, and dephasing noise in both single-qubit [Fig.~\ref{fig2}(a)] and two qubits [Fig.~\ref{fig2}(b)] systems through numerical simulations. The definitions of these three noise types and the noise level $\delta$ can be found in the Supplementary Materials~\cite{SI}.
In our simulations, for each noise type and noise level (each point in Fig.~\ref{fig2}), we generate 10 random input states, apply a certain noise channel, and measure these states on 10 random observables after mitigation. The observables are of the form $U_1ZU_1^{\dagger}$ for one qubit and $U_2Z\otimes ZU_2^{\dagger}$ for two qubits, where $Z$ represents the Pauli-z operator, and $U_1$ and $U_2$ are randomly chosen unitary operators for the respective qubit systems.
For each of 10 simulations (corresponding to one random input state and one random observable), we apply the CPTP map $\mathcal{E}_{\mathcal{N}}$ $N=1000$ times to estimate the mitigated expectation value.
In each trial, the measurement yields either $+1$ or $-1$, corresponding to the eigenvalues $+1$ or $(-1)$ of the observable, analogous to single-shot measurements in an experiment.
After collecting $N$ measurement outcomes, we post-process them by incorporating the measurement results of the ancilla qubit recorded in the classical register, multiplying by $\alpha(i)\gamma$, and multiplying by zero where the $r+1$-th Kraus operator is applied.
The mean value of the post-processed results serves as the mitigated expectation value.
To estimate the variance of the mitigated expectation value, we repeat this procedure 10 thousand times, amounting to a total of $10^7$ applications of $\mathcal{E}_{\mathcal{N}}$. The theoretical variance is determined using Eq.~(\ref{eq:var_N}). By averaging the variances across the 10 different input states and observables, we obtain each data point shown in Fig.~\ref{fig2}.

In Fig.~\ref{fig2}, we also present the theoretical prediction for the variance averaged over independently Haar-random input states and observables. With sufficiently many samples drawn from the Haar measure~\cite{Hayden2004,Gross2007,Scott_2008,Dankert2009}, the Haar-averaged variance is given by:
\begin{equation}
\begin{split}
\mathrm{Var}_{\text{Haar}}(A)
=\gamma \mathrm{Tr}[\mathcal{E}(I) \Sigma_{i=1}^r K_i^{\dagger}K_i]\frac{\mathrm{Tr}[A^2]}{d^2} -\frac{\mathrm{Tr}[A^2]+\mathrm{Tr}[A]^2}{d(d+1)}. \nonumber
\label{eq:var_ave}
\end{split}
\end{equation}
Here, the observable in each single measurement is randomized as $O=UAU^{\dagger}$, where $U$ is drawn from a Haar-random unitary ensemble. For unital channels, where $\mathcal{E}(I)=I$, the Haar-averaged variance simplifies to:
%such as the depolarizing and dephasing channels, =&(\frac{\lVert{\Lambda}\rVert_1}{d})^2\frac{\mathrm{Tr}[A^2]}{d} -\frac{\mathrm{Tr}[A^2]+\mathrm{Tr}[A]^2}{d(d+1)}\\
\begin{equation}
\begin{split}
\mathrm{Var}_{\text{Haar}}(A)
=&\gamma\lVert{\Lambda}\rVert_1\frac{\mathrm{Tr}[A^2]}{d^2} -\frac{\mathrm{Tr}[A^2]+\mathrm{Tr}[A]^2}{d(d+1)}\\
=&(\frac{\lVert{\Lambda}\rVert_1}{d})^2 -\frac{1}{d+1},
\label{eq:var_ave_unital}
\end{split}
\end{equation}
where the second line holds when operator A is a Pauli operator and $\mathcal{E}$ is a depolarizing or dephasing channel. $\lVert{\cdot}\rVert_1$ denotes the trace norm. Further details regarding the deviation of the averaged variance under Haar randomization can be found in the Supplementary Materials~\cite{SI}.

Bosonic modes are among the most foundational physical systems for performing quantum information processing tasks, such as quantum memory beyond the break-even point~\cite{Ofek2016,Ni2023Nature,Sivak2023}, quantum communication for building quantum networks~\cite{Axline2018,Burkhart2021}, quantum simulations~\cite{HU2018molecular,Wang2020prx}, quantum metrology~\cite{WangNC2019Heisenberg,wang2021quantumenhanced,Deng2024,Pan2025}. The dominant error of bosonic modes is photon loss error (see Supplementary Materials~\cite{SI} for its Kraus operator representation). Reducing the impact of photon loss is therefore crucial for enhancing the performance of bosonic-mode-based quantum information processing. Figure~\ref{fig3} shows the mitigation of photon loss errors using our method. Our simulations demonstrate sampling overhead [Fig.~\ref{fig3}(a)] and advantages in terms of Kraus rank for mitigating photon loss in a bosonic mode compared with the previous method~\cite{Regula2021,Jiang2021Quantum} [Fig.~\ref{fig3}(b)]. The simulation process follows the same methodology as in Fig.~\ref{fig2}, with observable $U_d Z_d U_d^{\dagger}$. Here, $U_d$ is a randomly chosen unitary operator for the bosonic mode and $Z_d=I_d-2\ket{d-1}\bra{d-1}$, which consists of the identity operator $I_d$ and the $d$-th Fock state $\ket{d-1}\bra{d-1}$ for the bosonic mode. According to Eq.~(\ref{eq:var_ave}), we also present the average variance for mitigating photon loss in Fig.~\ref{fig3}(a). As the system dimension increases, the number of Kraus operators in the CPTP map $\mathcal{E}_{\mathcal{N}}$ using our method scales linearly with $d+1$, where $d$ is both the system dimension and the Kraus rank of the HPTP map representing the matrix inverse of the photon loss error. In contrast, using the previous method, the Kraus ranks of two decomposed CPTP maps $\mathcal{E}_1$ and $\mathcal{E}_2$ with optimization are nearly full-rank ($d^2$). This makes the implementation impractical for large system dimensions.

%virtual quantum broadcasting~\cite{Parzygnat2024PRL}
%A CPTP channel is unital, meaning that it preserves the identity operator when applied, leaving it unchanged.  The inverse of the channel, if the channel is invertible,  is a HPTP channel~\cite{Jiang2021Quantum} and is also unital, due to $\mathcal{N}^{-1}\circ\mathcal{N}(I)=\mathcal{N}^{-1}(I)=I$.

%\section{Conclusion}
\smallskip{}
\noindent\textit{Conclusion.-} We have presented an efficient and fully constructive method for implementing arbitrary HPTP maps. By adapting binary tree-structured quantum circuits designed for implementing arbitrary CPTP maps, we introduce a post-processing at the end of the circuit to achieve the implementation of arbitrary HPTP maps. From a practical perspective, our method offers notable advantages. First, it requires near-optimal experimental resources. Specifically, it utilizes only a two-level ancilla qubit to assist in applying joint unitary operations and a classical register to record measurement outcomes of the ancilla qubit. The required dimension of the ancilla system is much smaller than in the previous method. Second, our method features a straightforward construction process for HPTP maps, eliminating the need for numerical optimization. Additionally, we show that the constructed CPTP maps corresponding to the desired HPTP maps have a Kraus rank of at most $r+1$, and the circuit depth scales as $\mathrm{log}_2(r+1)$ using a binary tree-structured quantum circuit. Beyond these practical benefits, on the theoretical side, we provide the variance of using our method for measuring arbitrary observables when implementing arbitrary HPTP maps on arbitrary input quantum states. Both the numerical and theoretical results highlight the efficiency and practicality of our method. We anticipate that this work will advance quantum control techniques and inspire broader exploration of non-completely-positive operations in quantum information processing.

\bigskip{}
\noindent \textbf{Acknowledgements}
\begin{acknowledgments}
\noindent This work was funded by the National Natural Science Foundation of China (Grants No. 92265210, 12550006, 92365301, 92165209, 92565301, 12547179 and 12574539), the Quantum Science and Technology-National Science and Technology Major Project (2021ZD0300200), and the Guangdong Provincial Quantum Science Strategic Initiative (Grant No.~GDZX2403008, GDZX2503001). X.Z. acknowledges support from the Chinese Ministry of Science and Technology (MOST) through grant 2023ZD0300600, from the Hong Kong Research Grant Council through grant 17310725, and from the State Key Laboratory of Quantum Information Technologies and Materials. This work is also supported by the Fundamental Research Funds for the Central Universities, the USTC Research Funds of the Double First-Class Initiative, the supercomputing system in the Supercomputing Center of USTC, and the USTC Center for Micro and Nanoscale Research and Fabrication.
\end{acknowledgments}

\end{document}